\documentclass[aps,reprint,floatfix,superscriptaddress]{revtex4-2}
\usepackage[english]{babel}
\usepackage[utf8]{inputenc}                                                            
\usepackage{epsfig}
\usepackage{amsmath}
\usepackage{units}
\usepackage[colorlinks]{hyperref}
\usepackage{enumerate}
\usepackage{bbold}
\usepackage{color}		  				

\begin{document}

\title{Chiral symmetry breaking through spontaneous dimerization in kagom\'e metals}
\author{Riccardo Ciola}
\affiliation{Institute for Theoretical Physics and Center for Extreme Matter and Emergent Phenomena,
Utrecht University, Princetonplein 5, 3584 CC Utrecht, The Netherlands}
\author{Kitinan Pongsangangan}
\affiliation{Institute for Theoretical Physics and Center for Extreme Matter and Emergent Phenomena,
Utrecht University, Princetonplein 5, 3584 CC Utrecht, The Netherlands}
\author{Ronny Thomale}
\affiliation{Institut f\"ur Theoretische Physik und Astrophysik,
Universit\"at W\"urzburg, 97074 W\"urzburg, Germany}
\author{Lars Fritz}
\affiliation{Institute for Theoretical Physics and Center for Extreme Matter and Emergent Phenomena,
Utrecht University, Princetonplein 5, 3584 CC Utrecht, The Netherlands}

\begin{abstract}
Due to an uprise in the variety of candidate compounds, kagome metals have recently gained significant attention. Among other features, kagome metals host Dirac cones as a key band structure feature away from half filling, and potentially yield an exceptionally large fine structure, beyond values found in other 2D Dirac materials such as graphene. We investigate the possibility of chiral symmetry breaking in kagome metals. Based on a heuristic lattice model, we determine the critical coupling strength and the ordering pattern by means of a Schwinger-Dyson mean-field analysis. As the leading instability we identify a dimerization pattern which spontaneously opens an excitation gap at the Dirac point and breaks the chiral symmetry. 
\end{abstract}
\maketitle

\section{Introduction}\label{sec:introduction}
Recent years have seen a plethora of condensed matter research activities in the field of Dirac- and Weyl-type systems, where effective Lorentz covariance is preserved due to the linear dispersion relation in the crystal. One of the first systems that gained massive attention is graphene \cite{CastroNeto2009}, but soon thereafter many more followed. 
Dirac systems exhibit a linear bandcrossing at isolated points in the Brillouin zone, which is locally protected by chiral and time reversal symmetry. As a result, Dirac systems for two spatial dimensions and higher show a semi-metallic density of states~\cite{CastroNeto2009,Armitage2018}. 
A discussion that predates graphene, and has its counterpart in high energy physics  \cite{Miransky1994}, centers around the possibility to spontaneously break the chiral symmetry in a two-dimensional layer of graphite, first formulated by Khveshchenko~\cite{Khveshchenko&Leal2004,Khveshchenko2009}. In contemporary condensed matter terms, this question translates into whether graphene at zero temperature is an (excitonic) insulator or not. The driving force behind the spontaneous gap formation in this scenario is Coulomb interaction whose strength is controlled by the dimensionless fine structure constant 
\begin{eqnarray}\label{Eq:alpha}
\alpha=\frac{e^2}{4\pi \epsilon_0 \epsilon_r v_F}\;.
\end{eqnarray}
Here, $e$ is the electric charge of the electron, $\epsilon_0$ the dielectric constant of vacuum, $\epsilon_r$ the relative permittivity of the medium, and $v_F$ the Fermi velocity of the electronic system. In typical condensed matter systems, $\alpha$ is a quantity that is below unity. 
\begin{figure}[h]
\includegraphics[width=0.4\textwidth]{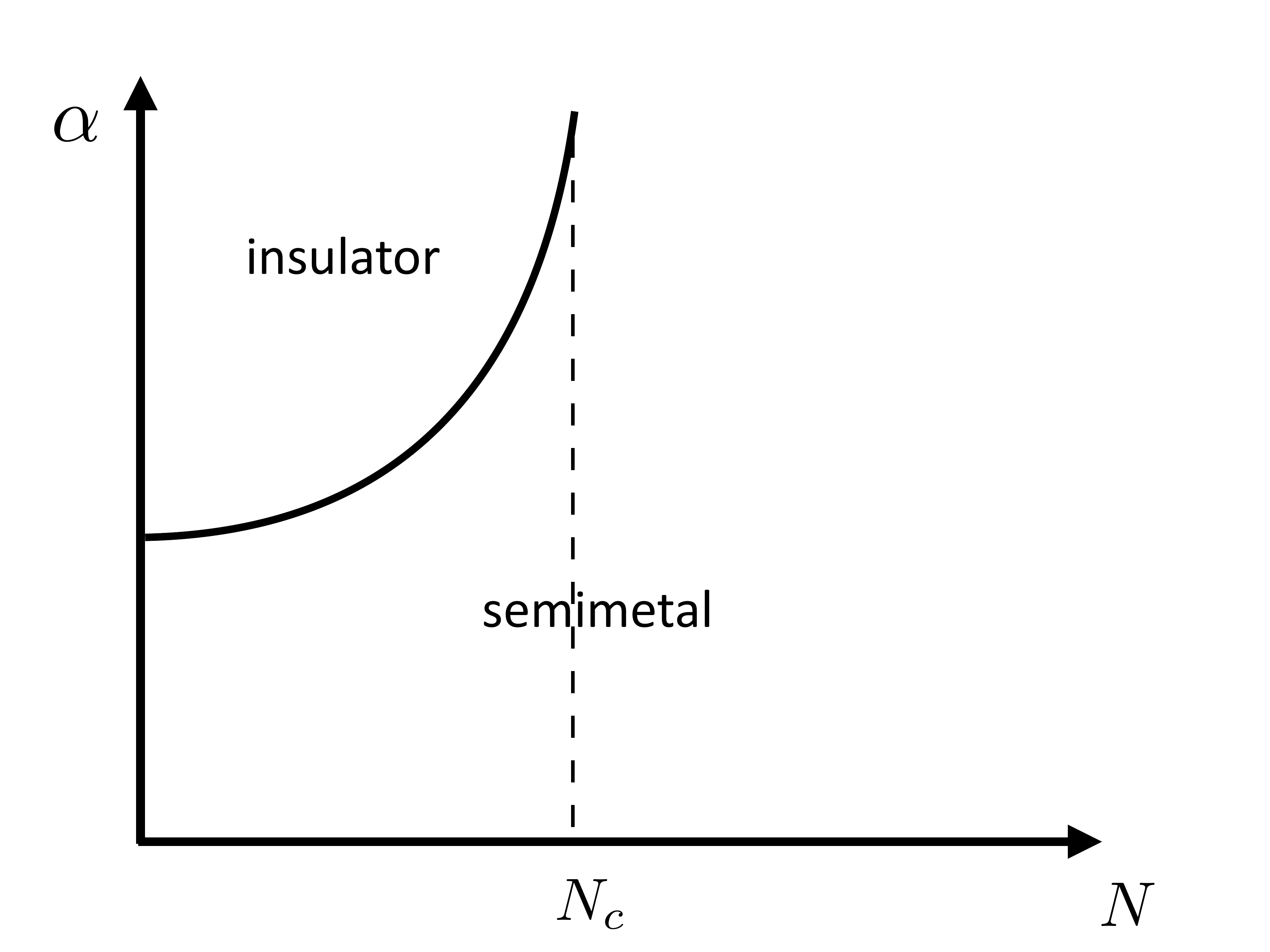}
\caption{Schematic phase diagram of two dimensional Dirac theories with long-range Coulomb interactions. The $x$-axis denotes the number of Dirac fermion flavors whereas the $y$-axis shows the value of the dimensionless coupling constant $\alpha$ (Eq.~\eqref{Eq:alpha}). For $N$ smaller than $N_c$, there exists a critical coupling strength $\alpha_c(N)$ where a quantum phase transition between a semimetal and an insulator takes place.}\label{fig:phasediagram}
\end{figure}

The problem of $N$ flavors of two-dimensional massless Dirac fermions with an effective speed $v_F$ coupled via long-range interactions has been investigated by a plethora of methods, ranging from renormalization group \cite{Son2007} to full lattice simulations using quantum Monte Carlo (for Dirac fermions this is possible without the sign problem due to the particle-hole symmetry of the system)~\cite{Ulybyshev2013}. There is a curious dichotomy in this field: perturbation theory based methods, including large-N approaches, suggest the irrelevance of Coulomb interaction irrespective of the value of the fine structure constant; on the other hand, strong coupling methods suggest the existence of a critical $\alpha_c$ beyond which there is spontaneous gap formation due to spontaneous chiral symmetry breaking, at least if $N$ is not too large. 
The finding of the strong coupling methods can be summarized in the schematic phase diagram shown in Fig.~\ref{fig:phasediagram}. The $x$-axis is labeled with $N$, which is the number of Dirac fermion flavors. In the case of graphene, this corresponds to $N=4$, which includes valley ($K$ and $K'$) and spin ($\uparrow$ and $\downarrow$). The $y$-axis denotes the strength of the Coulomb interaction, $\alpha$. There is a line where a quantum phase transition takes place. The critical coupling $\alpha_c$ associated with it goes to $\infty$ at $N_c$. This implies that there is a quantum phase transition from a semimetallic phase to an insulator at a critical coupling $\alpha_c$ as long as $N<N_c$. Depending on the calculational method, there is a large variation of $\alpha_c$ for a given $N$ \cite{Gamayun2010,Jing-RongWang2012,Carrington2016,Tupisyn2017}. 

In our work, we turn to kagom\'e metals as an intriguing instance of 2D Dirac materials, and their potential to show the interaction-driven phenomenon of chiral symmetry breaking. Kagom\'e metals have been investigated in a number of contexts. Descending from the established domain of Mott phases in kagom\'e layers as a host for quantum paramagnetism~\cite{RevModPhys.88.041002}, kagom\'e metals such as FeSn and Fe$_2$Sn$_3$ were found to exhibit itinerant magnetism and, in particular, anomalous Hall responses~\cite{ye-nature,doi:10.1063/1.5111792}. A host material for kagom\'e metals with a symmetry unbroken Fermi liquid parent state was most recently found in AV$_3$Sb$_5$ (A=K,Cs,Rb) featuring Vanadium kagom\'e nets~\cite{PhysRevMaterials.3.094407, Jiang2021}. At half filling, the generic kagom\'e band structure is close to (possibly multiple) van Hove singularities, and hence prone to exhibiting Fermi surface instabilities~\cite{PhysRevB.86.121105,Thomale2013,PhysRevB.87.115135}. While the degree of experimental exploration is still at an early stage, unconventional chiral charge order and superconductivity have already been theoretically ascribed to AV$_3$Sb$_5$~\cite{FENG2021,denner2021analysis,PhysRevB.104.035142,PhysRevB.104.045122,wu2021nature}, which might appear possibly along with an onset of nematic order~\cite{Jiang2021}. Reaching the Dirac cone filling in a kagom\'e metal, however, in general is a challenging task. Depending on the sign of the hybridization which dictates the flat kagome band to locate above (below) the dispersive bands, a filling of 1/3 (2/3) has to be reached. While as of yet one is still short of an experimental realization, Sc-substituted Herbertsmithite has been identified as a promising candidate material~\cite{mazin}. In the context of establishing the notion of turbulent electron hydrodynamics, Sc-Herbertsmithite has subsequently been the pivotal starting point to investigate the fine structure constant inherent to kagom\'e Dirac materials, and was shown that
it potentially has a very large fine structure constant, up to three times larger than graphene~\cite{DiSante2020}. This gives credence to the assertion that if one were to look for strong correlation phenomena of Dirac cones, kagom\'e Dirac materials appear as an eminently suited domain.

By employing a Schwinger-Dyson mean-field approach on the kagome lattice, we show that if the chemical potential of a kagom\'e metal is tuned to its Dirac point, there is an instability towards the spontaneous formation of a dimerization pattern (related patterns have previously been discussed in Refs.~\cite{Indergand2006,Guo2009}). This pattern breaks the sublattice symmetry and hence parity. As such, it opens a gap in the single-particle excitation spectrum on the mean-field level. We further find that the critical coupling for the dimerization formation is $\alpha_{\rm{cr}}\approx 1.22$, which from theoretical investigations for Sc-Herbertsmithite could be in realistic reach for tailored kagom\'e candidate materials.

Our paper is organized as follows. In Section ~\ref{Sec:Model}, we introduce our kagome model consisting of a generic tight-binding model supplemented by Coulomb interactions. We then proceed to the mean-field treatment of the model, and derive both the ordering pattern as well as the critical coupling in Section~\ref{Sec:Mean-Field}. In Section~\ref{Sec:Conclusion} we conclude that kagome metals promise to be among the most preferable systems to study spontaneous chiral symmetry breaking in Dirac materails.

\section{Model}\label{Sec:Model}

The starting point is the tight-binding model on the kagom\'e lattice. It has the generic form 
\begin{eqnarray}\label{eq:Hamiltonian}
H=t\sum_{<i,j>,\sigma}\left(c_{i\sigma}^\dagger c^{\phantom{\dagger}}_{j\sigma}+c_{j\sigma}^\dagger c^{\phantom{\dagger}}_{i\sigma} \right)\;.
\end{eqnarray}
The kagom\'e lattice has a three-site unit cell which we account for by introducing the three component vector $\psi_i=(A_i,B_i,C_i)^T$ which contains the wave-functions of one unit-cell. We furthermore introduce the three lattice vectors $\vec{a}_1=a(2,0)^T$, $\vec{a}_2=a(1,\sqrt{3})^T$, and $\vec{a}_3=a(-1,\sqrt{3})^T$ as well as the nearest neighbor vectors $\vec{\delta}_{AB}=a(1,0)^T$, $\vec{\delta}_{AC}=a/2(1,\sqrt{3})^T$, and $\vec{\delta}_{BC}=a/2(-1,\sqrt{3})^T$.
\begin{figure}
	\centering
	\includegraphics[width=0.48\textwidth]{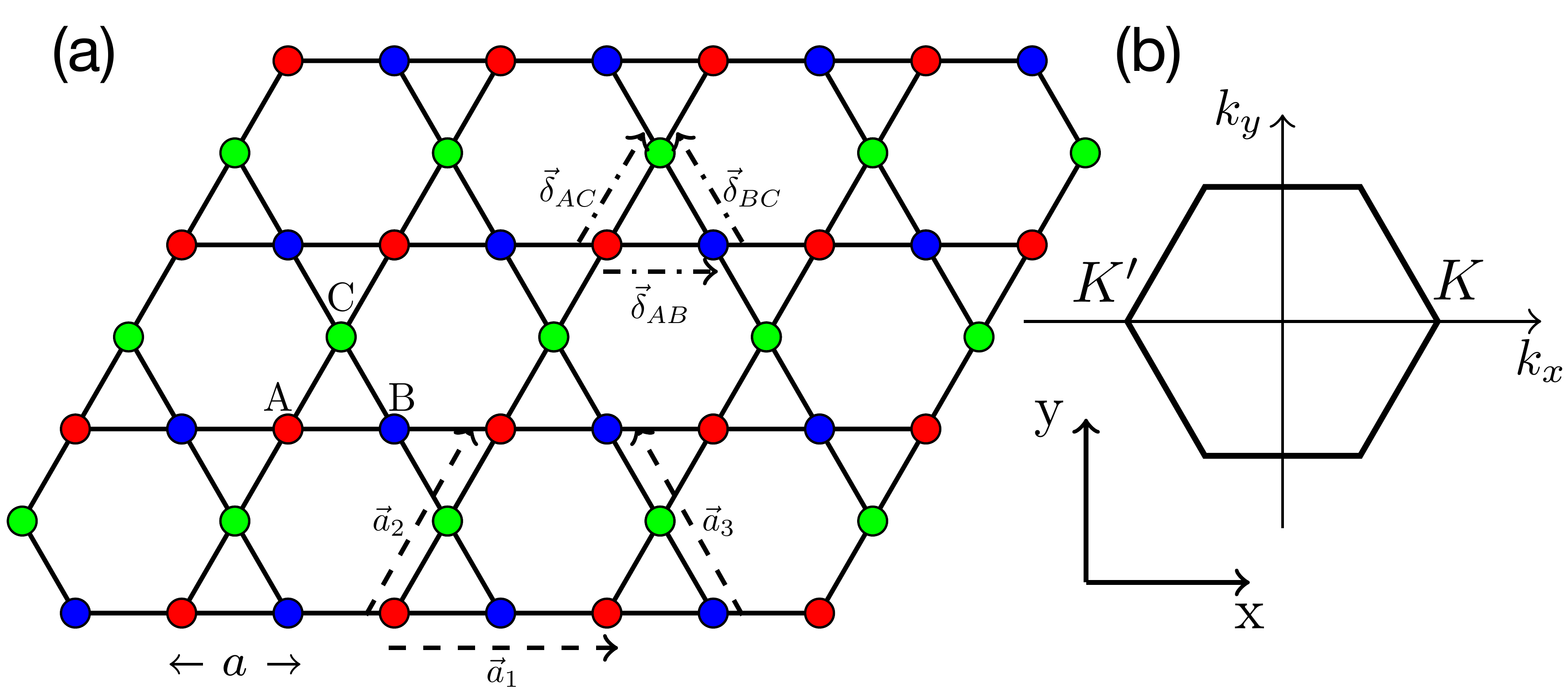}
	\caption{(a) Kagom\'e lattice with a three site unit cell. The sites within the unit cell are labelled $A$, $B$, and $C$. The vectors $\vec{a}_1$, $\vec{a}_2$, and $\vec{a}_3$ are the elementary lattice vectors defined in the main text. (b) The hexagonal first Brillouin zone of the kagom\'e lattice.}\label{fig:kagome}
\end{figure}

\begin{figure}[t]
\includegraphics[width=0.48\textwidth]{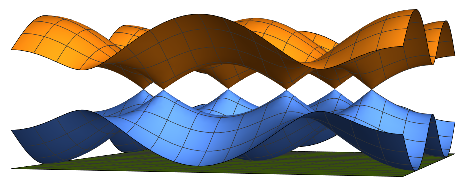}
\caption{Band structure of a kagom\'e nearest-neighbor tight-binding model. The band touching points at $2/3$-filling and their vicinities resemble the Dirac cones of graphene. }\label{Fig:BandStructure}
\end{figure}

Assuming periodic boundary conditions we can solve the Hamiltonian~\eqref{eq:Hamiltonian} via Fourier transform. The resulting Hamiltonian assumes the form
\begin{eqnarray}
H=\sum_\sigma \sum_{\rm{1st B.Z.}} \psi^\dagger_{\sigma,\vec{k}} \mathcal{H} (\vec{k}) \psi^{\phantom{\dagger}}_{\sigma,\vec{k}}
\end{eqnarray}
with the Bloch Hamiltonian given by
\begin{eqnarray}\label{Eq:Bloch}
\mathcal{H}(\vec{k})=2 t\left( \begin{array}{ccc}  0 & \cos \left(\vec{k}\cdot \vec{\delta}_{AB} \right) & \cos \left(\vec{k}\cdot \vec{\delta}_{AC} \right)  \\ \cos \left(\vec{k}\cdot \vec{\delta}_{AB} \right)  & 0 & \cos \left(\vec{k}\cdot \vec{\delta}_{BC} \right)  \\  \cos \left(\vec{k}\cdot \vec{\delta}_{AC} \right) & \cos \left(\vec{k}\cdot \vec{\delta}_{BC} \right)  & 0 \end{array}\right) \;. \nonumber \\
\end{eqnarray}
The spectrum of this system is given by 
\begin{eqnarray}
E_1 (\vec{k})&=& t \left(1+ \sqrt{3+2  \sum_{i}\cos \left (2\vec{k}\cdot \vec{\delta}_i\right)  }\right) \\ E_2 (\vec{k})&=& t \left(1- \sqrt{3+2  \sum_{i}\cos \left (2\vec{k}\cdot \vec{\delta}_i\right)  }\right) \\ E_3 (\vec{k})&=& -2t
\end{eqnarray}
with $\vec{\delta}_i=\vec{\delta}_{AB},\vec{\delta}_{AC},\vec{\delta}_{BC}$. The corresponding band structure is shown in Fig.~\ref{Fig:BandStructure}.
This Hamiltonian harbors a flat band at energy $-2t$ as well as a Dirac type band-structure reminiscent of graphene. The Dirac points are located at momenta $\vec{K}=(2\pi/3,0)^T$ and $\vec{K}'=(-2\pi/3,0)^T$. In the following we show how this Dirac structure emerges from the lattice Hamiltonian.

\subsection{Low Energy Expansion}
In graphene, it is sufficient to expand the Bloch Hamiltonian around one of the two independent Dirac point in order to obtain the linear Dirac cone structure that determines the massless electron dispersion. For the kagom\'e lattice, the tight-binding model, Eq.~\eqref{Eq:Bloch}, does not show the same simple structure and one has to explicitly decouple the flat band. If we expand the Bloch Hamiltonian up to the linear order in $\vec{K}+\vec{k}	$, we get
\begin{widetext}
\begin{eqnarray}
    \mathcal{H}(\vec{k}) \simeq -t\left( \begin{array}{ccc}
    0 & 1 + \sqrt{3}ak_x& -1 + \frac{\sqrt{3}}{2}ak_x + \frac{3}{2}ak_y\\
    1 + \sqrt{3}ak_x & 0 & -1 + \frac{\sqrt{3}}{2}ak_x - \frac{3}{2}ak_y\\
    -1 + \frac{\sqrt{3}}{2}ak_x + \frac{3}{2}ak_y&-1 + \frac{\sqrt{3}}{2}ak_x - \frac{3}{2}ak_y & 0
    \end{array} \right) \,.
\end{eqnarray}
\end{widetext}
We take a set of orthogonal eigenvectors of the linear Hamiltonian $\mathcal{H}(\vec{k} = 0)$ corresponding to the different bands to decouple the flat band and isolate the effective Dirac point theory. The proper transformation matrix is given by
\begin{eqnarray}
    U = \begin{bmatrix}
    \frac{1}{\sqrt{2}}\begin{pmatrix}
    0 \\ 1 \\ 1
    \end{pmatrix} \ , \
    \frac{1}{\sqrt{6}}\begin{pmatrix}
    2 \\ -1 \\ 1
    \end{pmatrix} \ , \
    \frac{1}{\sqrt{3}}\begin{pmatrix}
    -1 \\ -1 \\ 1
    \end{pmatrix}
    \end{bmatrix} \ .
\end{eqnarray}
which transforms $U\ h(\vec{k})\ U^{-1}=\mathcal{H}(\vec{k})$ with 
\begin{widetext}
\begin{eqnarray}
    h(\vec{k}) = t \left( \begin{array}{ccc}
    1-\frac{\sqrt{3}}{2}ak_x +\frac{3}{2}a k_y& -\frac{3}{2}ak_x-\frac{\sqrt{3}}{2}a k_y & \frac{3}{2\sqrt{2}}a k_x+\frac{\sqrt{3}}{2\sqrt{2}}a k_y\\
     -\frac{3}{2}ak_x-\frac{\sqrt{3}}{2}a k_y& 1+\frac{\sqrt{3}}{2}ak_x -\frac{3}{2}a k_y & \frac{\sqrt{3}}{2\sqrt{2}}a k_x+\frac{3}{2\sqrt{2}}a k_y\\
   \frac{3}{2\sqrt{2}}a k_x+\frac{\sqrt{3}}{2\sqrt{2}}a k_y & \frac{\sqrt{3}}{2\sqrt{2}}a k_x+\frac{3}{2\sqrt{2}}a k_y & -2
    \end{array} \right) \; .
\end{eqnarray}
\end{widetext}
In order to better identify the Dirac theory, we make the following coordinate transformation 
\begin{eqnarray}
k_1&=&\frac{1}{2}\left( k_x-\sqrt{3} k_y\right)\;, \nonumber \\
k_2&=&\frac{1}{2}\left( \sqrt{3}k_x+ k_y\right)\;.
\end{eqnarray}
which has a Jacobian of unity. We furthermore introduce $v_F=\sqrt{3}ta$ as the effective Fermi velocity. This allows to rewrite 
\begin{eqnarray}
    h(\vec{k}) = t \left( \begin{array}{ccc}
    t-v_F k_1 & -v_F k_2 & \frac{v_F}{\sqrt{2}}k_2\\
     -v_F k_2& t+v_F k_1 & \frac{v_F}{\sqrt{2}}k_1\\
 \frac{v_F}{\sqrt{2}}k_2 & \frac{v_F}{\sqrt{2}}k_1 & -2t
    \end{array} \right) \; .
\end{eqnarray}
The $2\times2$ matrix in the upper-left corner resembles the structure of the Dirac theory of graphene which we are going to show now. The other terms couple the effective Dirac theory to the flat band of energy $-2t$. To decouple the flat band, we integrate it out and derive an effective low energy theory. We do this perturbatively order by order, since the energy of the flat band is far removed from the Dirac point. For reasons of conciseness, we only sketch the procedure here: The leading order corrections are of the order $k_1^2/(3t)$, $k_2^2/(3t)$, and $k_1k_2/(3t)$ and can consequently safely be ignored in the linear limit. To conclude, to leading order we can ignore the rest and focus on the $2 \times 2$ submatrix, the upper left block:
\begin{eqnarray}
    \mathcal{H}^{\rm{Dirac}}(\vec{k}) &=&t\ \mathbb{1}+  v_F\left( k_1\sigma_z + k_2\sigma_x \right) \ .
    \label{HK-lin}
\end{eqnarray}
The effective dispersion of this Hamiltonian is given by the energy eigenvalues $E_{\pm}=t\pm v_F \sqrt{k_1^2+k_2^2}=t\pm v_F \sqrt{k_x^2+k_y^2}$, where the last step is easy to verify.

The linearized Hamiltonian~\eqref{HK-lin}, valid around $\vec{K}$, provides an excellent starting point for the discussion of potential mass terms. Excluding the possibility of hybridizing spin or valleys (for instance through disorder or short range interactions) for now, we concentrate on the two-dimensional matrix. The electron momenta are locked to the Pauli matrices $\sigma_x$ and $\sigma_z$, which leaves $\sigma_y$ as the 'free' Pauli matrix.  A term that cannot be eliminated by a momentum shift consequently is given by $ \Delta \sigma_y$. This leads to the dispersion $E_{\pm}=t\pm \sqrt{v_F^2 \vec{k}^2+\Delta^2}$ and consequently a gap of size $2|\Delta|$ at the Dirac points.

\subsection{Energy gap through dimerization}
 In a next step, we transform the mass term back to the full lattice problem which allows for a straightforward physical interpretation. Using the unitary matrix $U$, we transform the mass term back to the $3\times3$ lattice problem and end up with
\begin{equation}
    U \ \begin{pmatrix} 
    \Delta \sigma_y & \begin{matrix}
    0 \\
    0
    \end{matrix} \\
    \begin{matrix}
    0 & 0
    \end{matrix} & 0
    \end{pmatrix} \ U^{-1} = \frac{i\Delta}{\sqrt{3}} \begin{pmatrix}
    0 & 1 & 1 \\
    -1 & 0 & -1 \\
    -1 & 1 & 0
    \end{pmatrix} \ .
    \label{Eq:Mass}
\end{equation}

\begin{figure}
\includegraphics[width=0.4\textwidth]{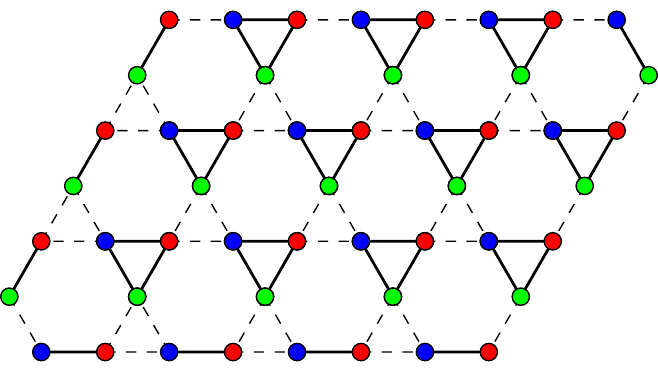}\caption{Real space representation of the dimerization pattern. Thick line corresponds to an enhanced hopping element whereas broken line corresponds to a reduced one, compare Eq.~\eqref{Kagome-dim}.}\label{Fig:DimPattern}
\end{figure}
This term corresponds to an inequivalence in the hopping between the sublattices inside a unit cell. It can therefore be identified with a dimerization pattern. In view of our later lattice analysis it is useful to compare it with a lattice version of a dimerization pattern~\cite{Guo2009,Indergand2006} which is shown in Fig.~\ref{Fig:DimPattern}.  After Fourier transformation, the dimerization matrix reads
\begin{eqnarray}
&&\delta \mathcal{H}(\vec{k})= \nonumber \\
   &&\frac{2 \Delta i}{3}  \left(  \begin{array}{ccc}
    0 & \sin(\vec{k}\cdot\vec{\delta}_{AB}) & \sin(\vec{k}\cdot\vec{\delta}_{AC})\\
    - \sin(\vec{k}\cdot\vec{\delta}_{AB}) & 0 & \sin(\vec{k}\cdot\vec{\delta}_{BC}) \\
    -\sin(\vec{k}\cdot\vec{\delta}_{AC}) & - \sin(\vec{k}\cdot\vec{\delta}_{BC}) & 0
    \end{array}  \right) \; . \nonumber \\
    \label{Kagome-dim}
\end{eqnarray}
At the Dirac point $\vec{K}$, we reproduce Eq.~\eqref{Eq:Mass} whereas at $\vec{K}'$ we obtain a relative sign (this is required by time-reversal symmetry, otherwise the system would have a finite Chern number and consequently be a Chern insulator). 
The total Bloch Hamiltonian reads $\mathcal{H}_{\rm{tot}}(\vec{k}) = \mathcal{H}(\vec{k}) + \delta \mathcal{H}(\vec{k})$ and has the spectrum
\begin{widetext}
\begin{eqnarray}
    E_{1,2}(\vec{k}) &=& t \left( 1 \pm \sqrt{3 \left(1 + \frac{2}{3}\sum_{j} \cos(2\vec{k}\cdot\vec{\delta}_j) \right)+\frac{2\Delta^2}{3t^2}\left(1 -\frac{1}{3} \sum_{j} \cos(2\vec{k}\cdot\vec{\delta}_j)\right) } \right)\;, \nonumber \\
    E_3(\vec{k}) &=& -2t\;.
\label{Energy-dim}
\end{eqnarray}
\end{widetext}
We observe that this dimerization process does not influence the flat band at all. 
\begin{figure}[t]
\includegraphics[width=0.48\textwidth]{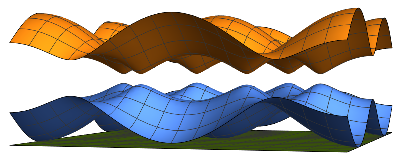}
\caption{Band structure of a kagom\'e nearest-neighbor tight-binding model with a dimerization term ($t=1$ and $\Delta=0.15$). The band touching points are removed in the presence of the dimerization term. }\label{Fig:BandStructureMass}
\end{figure}
To conclude: we found that if we introduce a mass function proportional to the dimerization matrix, Eq.~\eqref{Kagome-dim}, we will be able to open an energy gap at the Dirac point.

\subsection{Effective interaction}

\begin{center}
\begin{figure}
    \centerline{\includegraphics[width=0.45\textwidth]{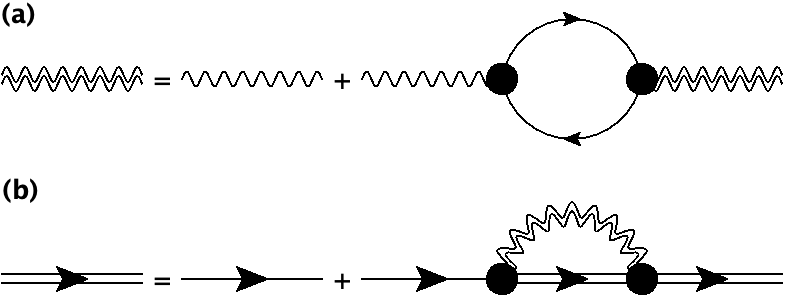}}    \caption{(a) Diagrammatic representation of the RPA potential equation. The double wiggly line is the renormalized Coulomb potential, whereas the single wiggly line is the bare Coulomb potential. The simple line is the fermion propagator and the closed bubble the polarization function. (b) The self-consistent Schwinger Dyson equation. The double line is the full fermion propagator.}
    \label{fig-Vrpa}
\end{figure} 
\end{center}

In non-relativistic electronic system ($v_F\ll c$ with $c$ being the speed of light), Coulomb interaction assumes the form
\begin{eqnarray}
&&\frac{e^2}{8\pi \epsilon_0 \epsilon_r} \int d^dr d^dr' \rho(\vec{r})\frac{1}{|\vec{r}-\vec{r}'|}\rho(\vec{r}') \nonumber \\ && = \frac{\alpha v_F}{2}\int d^dr d^dr' \rho(\vec{r})\frac{1}{|\vec{r}-\vec{r}'|}\rho(\vec{r}')\;,
\end{eqnarray}
where $\rho(\vec{r})$ is the electron density at site $\vec{r}$. The measure for the strength of the electronic interaction is the dimensionless fine-structure constant $\alpha$ which was introduced in Eq.~\eqref{Eq:alpha}. For typical Dirac systems it is a quantity $\mathcal{O}(1)$, since $v_F/c \approx 10^{-2}-10^{-3}$ and $\epsilon_r \approx 2 -10$.
Typically, in finite density electronic systems, Coulomb interaction is dynamically screened through particle-hole excitations. Technically, this can described by the random phase approximation (RPA). RPA resums the whole diagrammatic series shown in Fig.~\ref{fig-Vrpa} (a) and can formally be justified in the limit of a large number of fermion flavors. The double wiggly (single) line denotes the effective RPA potential (unscreened potential), $V^{\rm{RPA}}\left(\vec{k}, \omega\right)$, whereas the straight directed line is the fermion propagator. 
The effective RPA potential assumes the form 
\begin{eqnarray}
        V^{\rm{RPA}}\left(\vec{k}, \omega\right) =  \frac{1}{V^{-1}_C(\vec{k}) - \Pi\left(\vec{k}, \omega\right)}  \; ,
    \label{Vrpa}
\end{eqnarray}
where $\Pi\left(\vec{k}, \omega\right)$ is the polarization function and the bare Coulomb interaction in two dimensions is given by
\begin{eqnarray}
V_C(\vec{k})=\frac{e^2}{2\epsilon}\frac{1}{|\vec{k}|}\;.
\end{eqnarray}

The polarization function at zero temperature can be expressed in terms of the fermion Green function according to
\begin{eqnarray}\label{Eq:Pol}
\Pi(\vec{k},\omega)=-2\int \frac{d\nu}{2\pi}\int \frac{d^2q}{(2\pi)^2} \; {\rm{tr}}\; G_0(\vec{k}+\vec{q},\omega+\nu)G_0(\vec{q},\nu)\;.\nonumber \\
\end{eqnarray}
In the above expression, the factor of two is related to the spin degree of freedom.
We use the Green function of the fermions in Eq.~\eqref{Eq:Pol} whose inverse is given by
\begin{eqnarray}
    G^{-1}_0(\vec{k}, \omega_n) = -i\omega_n  +\mathcal{H}(\vec{k}) \; ,
    \label{G0inv-Kagome}
\end{eqnarray}

where $\mathcal{H}(\vec{k})$ is the Bloch Hamiltonian of the kagom\'e system in absence of dimerization and $\omega_n=(2n+1)\pi T$ is the fermionic Matsubara frequency with $n$ being an integer. 

\begin{figure}[!htbp]
    \centerline{\includegraphics[width=0.48\textwidth]{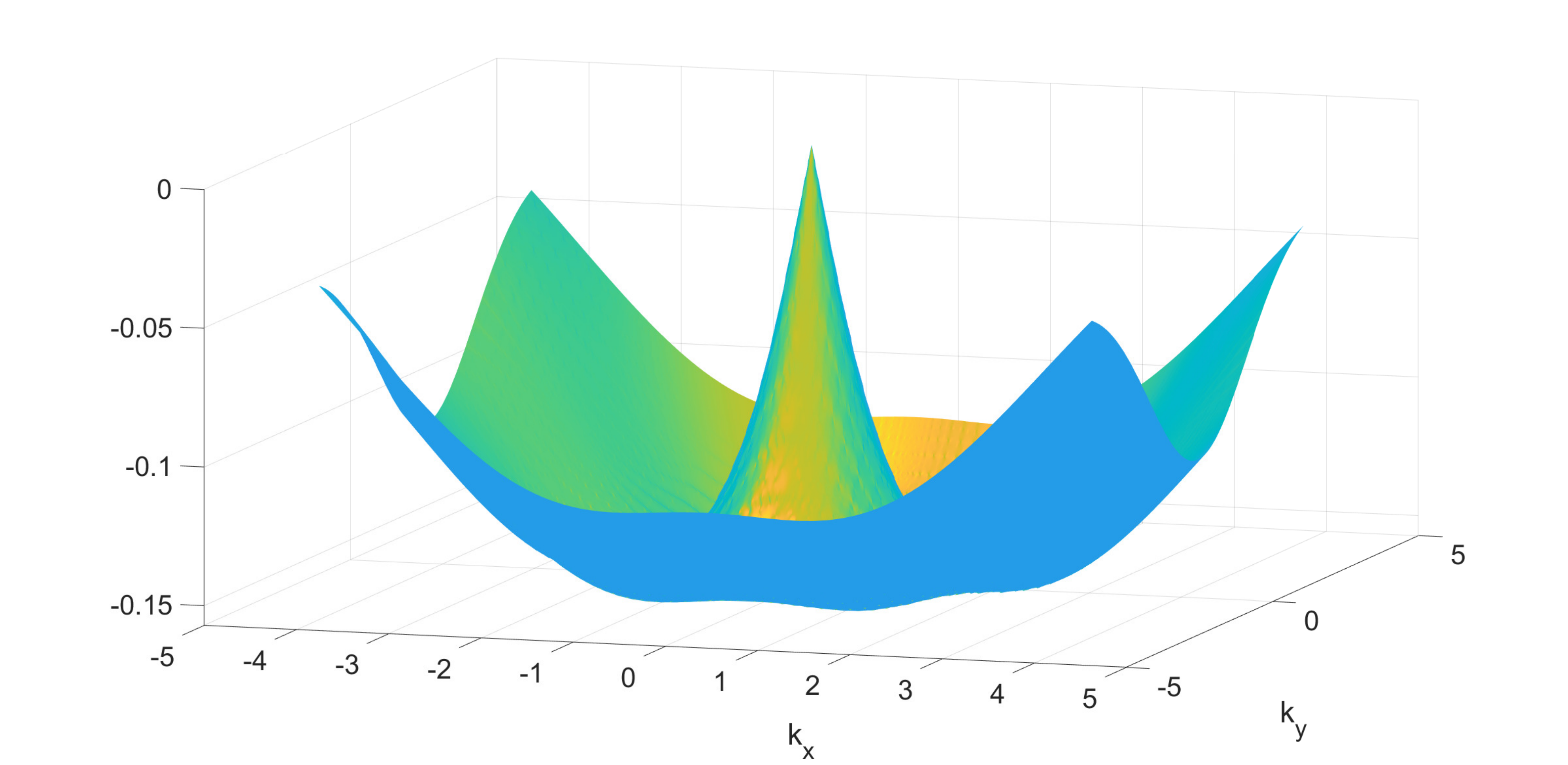}}    \caption{Polarization function for the Kagome lattice within the on-shell condition. }
    \label{fig:polarization-kagome}
\end{figure}

In pure two-dimensional Dirac systems at zero temperature, it is well known that the polarization function in terms of Matsubara frequencies follows
\begin{eqnarray}\label{Eq:DiracPol}
\Pi_{\rm{2D-Dirac}}(\vec{k},\omega) \propto \frac{\vec{k}^2}{\sqrt{v_F^2 \vec{k}^2+\omega^2}}\;,
\end{eqnarray}
which shows the absence of screening. 
We could not find a closed analytical expression for Eq.~\eqref{Eq:Pol} for the full three band system, so we had to resort to a numerical solution instead. In order to simplify the representation and also in view of the later treatment, we use the on-shell condition, meaning $\omega=v_F k$, which is valid for low values of $\omega$. The result of the numerical integration is shown in Fig. \ref{fig:polarization-kagome}. In the respective plot, the chemical potential is tuned to the Dirac point. Importantly, it shows the characteristic $\propto |\vec{k}|/v_F$ at low values of $|\vec{k}|$ behavior also observed in the two-dimensional Dirac theory, see Eq.~\eqref{Eq:DiracPol}.  We checked that the slope agrees with the inverse value of the effective Fermi velocity relevant for the lattice system. 

\section{The Schwinger-Dyson equation}\label{Sec:Mean-Field}

\subsection{Meanfield analysis}
The Schwinger-Dyson mean-field technique is a self-consistent technique which allows to estimate the critical coupling of a (quantum) phase transition. On a formal level, it can be derived from a Hubbard-Stratonovich decoupling of the Coulomb interaction in the respective instability channel, in this case the dimerization. It leads to a self-consistency equation of the type
\begin{equation}
    \delta \mathcal{H}(\textbf{k})  = \int \frac{d^2p}{(2\pi)^2}\frac{d\nu}{2\pi} \ V^{\rm{RPA}}(\vec{k}+\vec{p}) G(\vec{p},\nu)\; ,
    \label{Kagome-SD1}
\end{equation}
which is diagrammatically shown in Fig.~\ref{fig-Vrpa} (b). In this expression,
\begin{eqnarray}
    G^{-1}(\vec{k}, \omega) = -i\omega  +\mathcal{H}(\vec{k})+\delta \mathcal{H}(\vec{k}) \; ,
    \label{Ginv-Kagome}
\end{eqnarray}
is the Green function of the system taking into account the presence of the dimerization function. In Eq.~\eqref{Kagome-SD1}, we have employed the on-shell condition for the interaction potential $V^{\rm{RPA}}$, {\it i.e.}, it is only a function of $\vec{k}+\vec{p}$. The poles in the $\nu$-integral reside at 
\begin{eqnarray}
\nu_{1,2}(\vec{p},\Delta)=-i E_{1,2}(\vec{p})\;,
\end{eqnarray}
see Eq.~\eqref{Energy-dim}. Interestingly, the flat band vanishes exactly from the gap equation, leaving only the Dirac cones. This is to be expected since according to Eq.~\eqref{Kagome-dim} a finite dimerization leaves the flat band inert.
We can now shift the frequency according to the chemical potential $t$, namely $\nu' \rightarrow \nu + it$. In this way the poles become symmetric around the real axis, {\it i.e.}, $\nu'_{\pm} = \nu_{\pm}-it$, and we can integrate the Schwinger-Dyson Eq.~\eqref{Kagome-SD1} using the residue theorem. The self-consistent equation for the gap becomes
\begin{eqnarray}
   \Delta (\vec{k}) = \int_{BZ} \frac{d\vec{p}}{(2\pi)^2} \  \ \frac{V^{\rm{RPA}}(\vec{k}+\vec{p}) \Delta(\vec{p})}{2i\nu'_+\left(\vec{p}, \Delta(\vec{p})\right)} \ \frac{\sin(\vec{p}\cdot\vec{\delta}_j)}{\sin(\vec{k}\cdot\vec{\delta}_j)} \ .\nonumber \\
    \label{Kagome-DS2}
\end{eqnarray}
It is important to note that compared to Eq.~\eqref{Kagome-dim} we allow for a mean-field parameter $\Delta$ that is a function of $\vec{k}$, {\it i.e.}, $\Delta(\vec{k})$.
This expression is independent of the specific direction j. Without loss of generality we choose $\vec{\delta}_{AB}$ leading to a gap function $\Delta(\vec{p})$ that is symmetric in all the Dirac points, {\it{i.e.}}, symmetric under a rotation of the Brillouin zone of $2\pi/3$ radians, in agreement with the symmetry of the underlying lattice. 
One comment about Eq.~\eqref{Kagome-DS2} is in order here: we are mostly interested in the gap function $\Delta$ for $\vec{k} \approx \vec{K}$, meaning at the Dirac point. Since $V^{\rm{RPA}}(\vec{k}+\vec{p})$ gives the maximal contribution at $\vec{p}=-\vec{k}$, this requires $\vec{p}\approx -\vec{K}$. This implies that the self-consistency equation couples opposite Dirac points. In that situation, $\sin(\vec{p} \cdot \vec{\delta}_j)/\sin(\vec{k}\cdot \vec{\delta}_j)\simeq -1$. This can be used to show that it is justified to carry out the Schwinger-Dyson mean-field treatment in the effective Dirac low-energy setting, as one should expect.

It is worth noting here that we calculated the RPA interaction, $V^{\rm{RPA}}$, using the bare propagator, $G_0$. While this results in a slightly inaccurate estimate of the gap term in the ordered phase, it does not influence the location of the critical point, which is our focus.

\subsection{Spontaneous dimerization}

\begin{figure}[!htbp]
    \centerline{\includegraphics[width=0.48\textwidth]{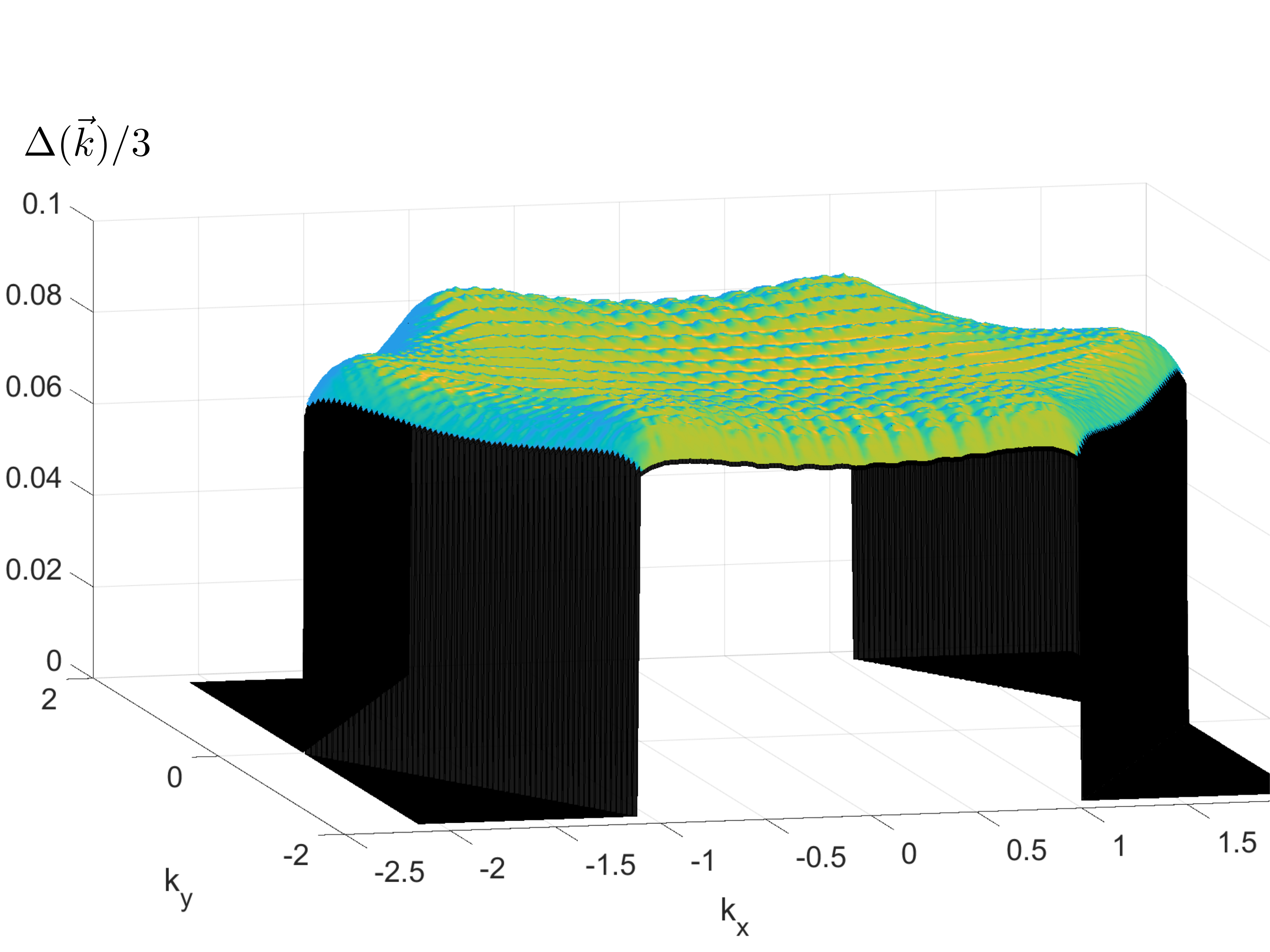}}   
     \caption{Numerical solution of the gap equation with coupling parameter $\alpha = 2.0$.}
    \label{fig:mass-Kagome}
\end{figure}

\begin{figure}[!htbp]    \centerline{\includegraphics[width=0.48\textwidth]{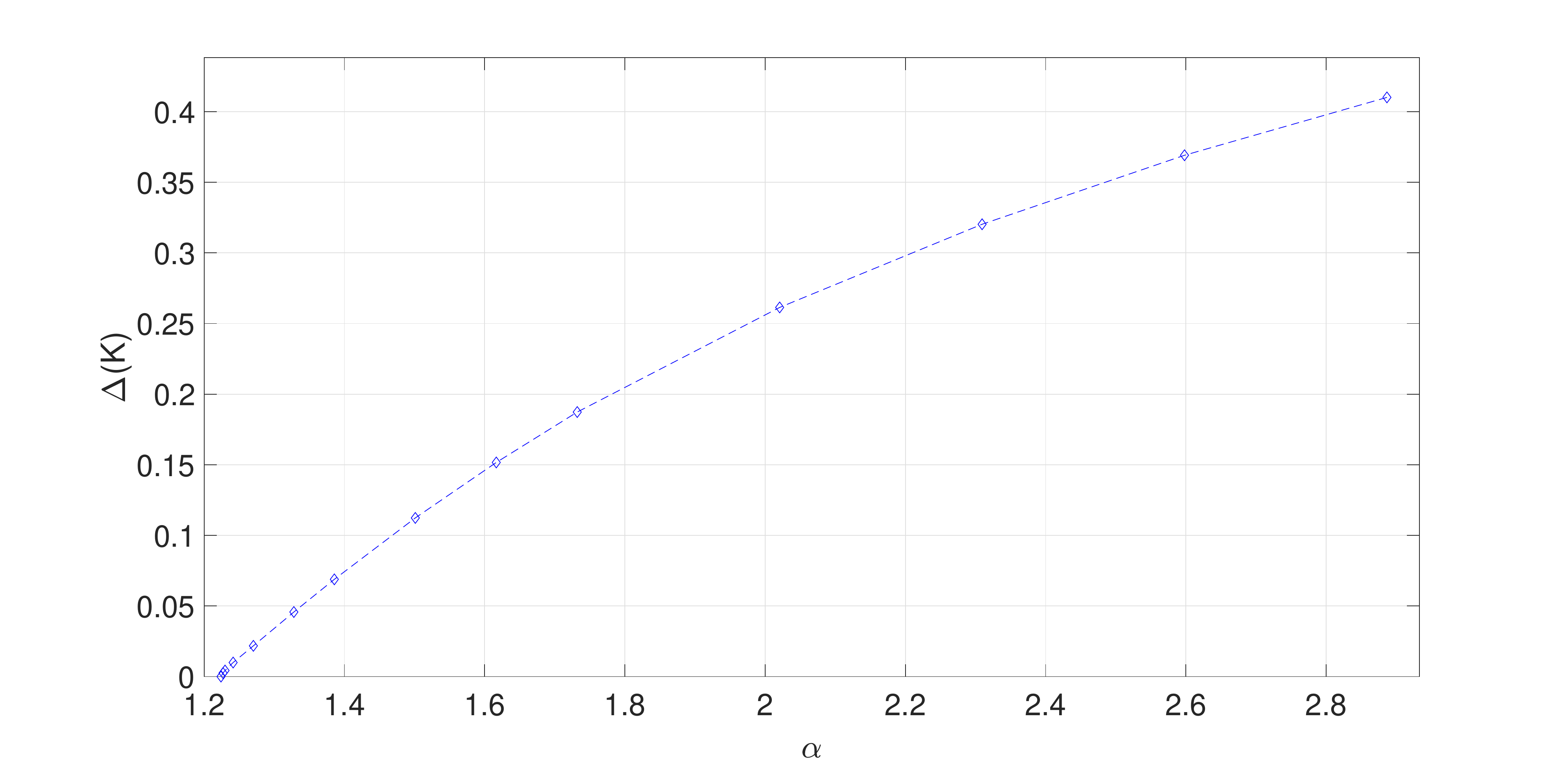}}    \caption{Mass gap at the Dirac point as a function of the coupling parameter $\alpha$. We find that there is a spontaneous gap formation for values of $\alpha >  1.224$}
    \label{fig:phase-diagram-Kagome}
\end{figure}

The gap equation, Eq.~\eqref{Kagome-DS2}, can now be solved numerically for different values of the coupling parameter $\alpha$. On a technical level, we use Gaussian quadrature for the integration and find the solution through an iteration of the gap equation, which stops when convergence is reached. A representative example of the gap solution profile is shown in Fig. \ref{fig:mass-Kagome} for a value of $\alpha=2.0$. \\
\\
We have determined the phase diagram of this semimetal-insulator transition as a function of $\alpha$. It shows a transition from a gapless semi-metallic phase to a gapped insulating phase. The value of the gap as a function of $\alpha$ is shown in Fig. \ref{fig:phase-diagram-Kagome} which is the main result of this paper.
\\
The critical coupling parameter is given by $\alpha_c \approx 1.224$: for $\alpha>\alpha_c$ there is spontaneous mass generation, whereas for $\alpha<\alpha_c$ the systems remains semi-metallic. It should be noted here, that this critical value is in good agreement with the value predicted from a simple two-component Dirac theory.

\section{Conclusion}\label{Sec:Conclusion}

Kagom\'e  metals offer an exciting new platform to observe many-body physics, mostly due to the existence of van Hove singularities. In this work we have focused on the Dirac metal structure also present in kagom\'e systems:  at $2/3$-filling, a kagom\'e metal such as Sc-Herbertsmithite, is expected to show physics characteristic of a Dirac metal. We have found that in that situation one can potentially observe the phenomenon of chiral symmetry breaking of Dirac fermions.This dynamic instability can be driven by long-range Coulomb interactions and on the microscopic level it corresponds to the formation of a dimerization pattern that opens a gap at the Dirac points. This instability occurs for values of $\alpha$ greater than $\alpha_c \approx 1.224$. It is the analogue of the excitonic insulator which was hypothesized to exist in suspended graphene. This has not been observed in experiments to date and a possible explanation is that realistic values of $\alpha$ are either too small ($\mathcal{O}(1)$) or the gap is too small to be detected. Given that e.g. recent estimates for the fine structure constant in Sc-Herbertsmithite suggest $\alpha \approx 2.9$ leads us to conclude that kagom\'e Dirac semi-metals constitute a promising system to look for spontaneous chiral symmetry breaking of Dirac fermions.

\section*{Acknowledgements}
The authors thank Johanna Erdmenger and Henk Stoof for discussions.
This work is part of the D-ITP consortium,
a program of the Netherlands Organisation for Scientific Research (NWO) that is funded by the Dutch Ministry of Education, Culture and Science (OCW).
R.T. is supported by the Deutsche Forschungsgemeinschaft (DFG, German Research Foundation) through Project-ID 258499086-SFB 1170 and the W\"urzburg-Dresden Cluster of Excellence on Complexity and Topology in Quantum Matter – ct.qmat Project-ID 390858490-EXC 2147.

\bibliographystyle{revtex4-2}

\begin{thebibliography}{28}
	\expandafter\ifx\csname natexlab\endcsname\relax\def\natexlab#1{#1}\fi
	\expandafter\ifx\csname bibnamefont\endcsname\relax
	\def\bibnamefont#1{#1}\fi
	\expandafter\ifx\csname bibfnamefont\endcsname\relax
	\def\bibfnamefont#1{#1}\fi
	\expandafter\ifx\csname citenamefont\endcsname\relax
	\def\citenamefont#1{#1}\fi
	\expandafter\ifx\csname url\endcsname\relax
	\def\url#1{\texttt{#1}}\fi
	\expandafter\ifx\csname urlprefix\endcsname\relax\def\urlprefix{URL }\fi
	\providecommand{\bibinfo}[2]{#2}
	\providecommand{\eprint}[2][]{\url{#2}}
	
	\bibitem[{\citenamefont{Castro~Neto et~al.}(2009)\citenamefont{Castro~Neto,
			Guinea, Peres, Novoselov, and Geim}}]{CastroNeto2009}
	\bibinfo{author}{\bibfnamefont{A.~H.} \bibnamefont{Castro~Neto}},
	\bibinfo{author}{\bibfnamefont{F.}~\bibnamefont{Guinea}},
	\bibinfo{author}{\bibfnamefont{N.~M.~R.} \bibnamefont{Peres}},
	\bibinfo{author}{\bibfnamefont{K.~S.} \bibnamefont{Novoselov}},
	\bibnamefont{and} \bibinfo{author}{\bibfnamefont{A.~K.} \bibnamefont{Geim}},
	\bibinfo{journal}{Rev. Mod. Phys.} \textbf{\bibinfo{volume}{81}},
	\bibinfo{pages}{109} (\bibinfo{year}{2009}),
	\urlprefix\url{https://link.aps.org/doi/10.1103/RevModPhys.81.109}.
	
	\bibitem[{\citenamefont{Armitage et~al.}(2018)\citenamefont{Armitage, Mele, and
			Vishwanath}}]{Armitage2018}
	\bibinfo{author}{\bibfnamefont{N.~P.} \bibnamefont{Armitage}},
	\bibinfo{author}{\bibfnamefont{E.~J.} \bibnamefont{Mele}}, \bibnamefont{and}
	\bibinfo{author}{\bibfnamefont{A.}~\bibnamefont{Vishwanath}},
	\bibinfo{journal}{Rev. Mod. Phys.} \textbf{\bibinfo{volume}{90}},
	\bibinfo{pages}{015001} (\bibinfo{year}{2018}),
	\urlprefix\url{https://link.aps.org/doi/10.1103/RevModPhys.90.015001}.
	
	\bibitem[{\citenamefont{Miransky}(1994)}]{Miransky1994}
	\bibinfo{author}{\bibfnamefont{V.~A.} \bibnamefont{Miransky}},
	\emph{\bibinfo{title}{Dynamical Symmetry Breaking in Quantum Field Theories}}
	(\bibinfo{publisher}{WORLD SCIENTIFIC}, \bibinfo{year}{1994}),
	\eprint{https://www.worldscientific.com/doi/pdf/10.1142/2170},
	\urlprefix\url{https://www.worldscientific.com/doi/abs/10.1142/2170}.
	
	\bibitem[{\citenamefont{Khveshchenko and Leal}(2004)}]{Khveshchenko&Leal2004}
	\bibinfo{author}{\bibfnamefont{D.}~\bibnamefont{Khveshchenko}}
	\bibnamefont{and} \bibinfo{author}{\bibfnamefont{H.}~\bibnamefont{Leal}},
	\bibinfo{journal}{Nuclear Physics B} \textbf{\bibinfo{volume}{687}},
	\bibinfo{pages}{323} (\bibinfo{year}{2004}).
	
	\bibitem[{\citenamefont{Khveshchenko}(2009)}]{Khveshchenko2009}
	\bibinfo{author}{\bibfnamefont{D.}~\bibnamefont{Khveshchenko}},
	\bibinfo{journal}{J. Phys.: Condens. Matter} \textbf{\bibinfo{volume}{21}}
	(\bibinfo{year}{2009}).
	
	\bibitem[{\citenamefont{Son}(2007)}]{Son2007}
	\bibinfo{author}{\bibfnamefont{D.~T.} \bibnamefont{Son}},
	\bibinfo{journal}{Phys. Rev. B} \textbf{\bibinfo{volume}{75}},
	\bibinfo{pages}{235423} (\bibinfo{year}{2007}),
	\urlprefix\url{https://link.aps.org/doi/10.1103/PhysRevB.75.235423}.
	
	\bibitem[{\citenamefont{Ulybyshev et~al.}(2013)\citenamefont{Ulybyshev,
			Buividovich, Katsnelson, and Polikarpov}}]{Ulybyshev2013}
	\bibinfo{author}{\bibfnamefont{M.~V.} \bibnamefont{Ulybyshev}},
	\bibinfo{author}{\bibfnamefont{P.~V.} \bibnamefont{Buividovich}},
	\bibinfo{author}{\bibfnamefont{M.~I.} \bibnamefont{Katsnelson}},
	\bibnamefont{and} \bibinfo{author}{\bibfnamefont{M.~I.}
		\bibnamefont{Polikarpov}}, \bibinfo{journal}{Phys. Rev. Lett.}
	\textbf{\bibinfo{volume}{111}}, \bibinfo{pages}{056801}
	(\bibinfo{year}{2013}),
	\urlprefix\url{https://link.aps.org/doi/10.1103/PhysRevLett.111.056801}.
	
	\bibitem[{\citenamefont{Gamayun et~al.}(2010)\citenamefont{Gamayun, Gorbar, and
			Gusynin}}]{Gamayun2010}
	\bibinfo{author}{\bibfnamefont{O.~V.} \bibnamefont{Gamayun}},
	\bibinfo{author}{\bibfnamefont{E.~V.} \bibnamefont{Gorbar}},
	\bibnamefont{and} \bibinfo{author}{\bibfnamefont{V.~P.}
		\bibnamefont{Gusynin}}, \bibinfo{journal}{Phys. Rev. B}
	\textbf{\bibinfo{volume}{81}}, \bibinfo{pages}{075429}
	(\bibinfo{year}{2010}),
	\urlprefix\url{https://link.aps.org/doi/10.1103/PhysRevB.81.075429}.
	
	\bibitem[{\citenamefont{Wang and Liu}(2012)}]{Jing-RongWang2012}
	\bibinfo{author}{\bibfnamefont{J.-R.} \bibnamefont{Wang}} \bibnamefont{and}
	\bibinfo{author}{\bibfnamefont{G.-Z.} \bibnamefont{Liu}},
	\bibinfo{journal}{New Journal of Physics} \textbf{\bibinfo{volume}{14}},
	\bibinfo{pages}{043036} (\bibinfo{year}{2012}),
	\urlprefix\url{https://doi.org/10.1088/1367-2630/14/4/043036}.
	
	\bibitem[{\citenamefont{Carrington et~al.}(2016)\citenamefont{Carrington,
			Fischer, von Smekal, and Thoma}}]{Carrington2016}
	\bibinfo{author}{\bibfnamefont{M.~E.} \bibnamefont{Carrington}},
	\bibinfo{author}{\bibfnamefont{C.~S.} \bibnamefont{Fischer}},
	\bibinfo{author}{\bibfnamefont{L.}~\bibnamefont{von Smekal}},
	\bibnamefont{and} \bibinfo{author}{\bibfnamefont{M.~H.} \bibnamefont{Thoma}},
	\bibinfo{journal}{Phys. Rev. B} \textbf{\bibinfo{volume}{94}},
	\bibinfo{pages}{125102} (\bibinfo{year}{2016}),
	\urlprefix\url{https://link.aps.org/doi/10.1103/PhysRevB.94.125102}.
	
	\bibitem[{\citenamefont{Tupitsyn and Prokof'ev}(2017)}]{Tupisyn2017}
	\bibinfo{author}{\bibfnamefont{I.~S.} \bibnamefont{Tupitsyn}} \bibnamefont{and}
	\bibinfo{author}{\bibfnamefont{N.~V.} \bibnamefont{Prokof'ev}},
	\bibinfo{journal}{Phys. Rev. Lett.} \textbf{\bibinfo{volume}{118}},
	\bibinfo{pages}{026403} (\bibinfo{year}{2017}),
	\urlprefix\url{https://link.aps.org/doi/10.1103/PhysRevLett.118.026403}.
	
	\bibitem[{\citenamefont{Norman}(2016)}]{RevModPhys.88.041002}
	\bibinfo{author}{\bibfnamefont{M.~R.} \bibnamefont{Norman}},
	\bibinfo{journal}{Rev. Mod. Phys.} \textbf{\bibinfo{volume}{88}},
	\bibinfo{pages}{041002} (\bibinfo{year}{2016}),
	\urlprefix\url{https://link.aps.org/doi/10.1103/RevModPhys.88.041002}.
	
	\bibitem[{\citenamefont{Ye et~al.}(2018)\citenamefont{Ye, Kang, Liu, von Cube,
			Wicker, Suzuki, Jozwiak, Bostwick, Rotenberg, Bell et~al.}}]{ye-nature}
	\bibinfo{author}{\bibfnamefont{L.}~\bibnamefont{Ye}},
	\bibinfo{author}{\bibfnamefont{M.}~\bibnamefont{Kang}},
	\bibinfo{author}{\bibfnamefont{J.}~\bibnamefont{Liu}},
	\bibinfo{author}{\bibfnamefont{F.}~\bibnamefont{von Cube}},
	\bibinfo{author}{\bibfnamefont{C.~R.} \bibnamefont{Wicker}},
	\bibinfo{author}{\bibfnamefont{T.}~\bibnamefont{Suzuki}},
	\bibinfo{author}{\bibfnamefont{C.}~\bibnamefont{Jozwiak}},
	\bibinfo{author}{\bibfnamefont{A.}~\bibnamefont{Bostwick}},
	\bibinfo{author}{\bibfnamefont{E.}~\bibnamefont{Rotenberg}},
	\bibinfo{author}{\bibfnamefont{D.~C.} \bibnamefont{Bell}},
	\bibnamefont{et~al.}, \bibinfo{journal}{Nature}
	\textbf{\bibinfo{volume}{555}}, \bibinfo{pages}{638} (\bibinfo{year}{2018}).
	
	\bibitem[{\citenamefont{Inoue et~al.}(2019)\citenamefont{Inoue, Han, Ye,
			Suzuki, and Checkelsky}}]{doi:10.1063/1.5111792}
	\bibinfo{author}{\bibfnamefont{H.}~\bibnamefont{Inoue}},
	\bibinfo{author}{\bibfnamefont{M.}~\bibnamefont{Han}},
	\bibinfo{author}{\bibfnamefont{L.}~\bibnamefont{Ye}},
	\bibinfo{author}{\bibfnamefont{T.}~\bibnamefont{Suzuki}}, \bibnamefont{and}
	\bibinfo{author}{\bibfnamefont{J.~G.} \bibnamefont{Checkelsky}},
	\bibinfo{journal}{Applied Physics Letters} \textbf{\bibinfo{volume}{115}},
	\bibinfo{pages}{072403} (\bibinfo{year}{2019}).
	
	\bibitem[{\citenamefont{Ortiz et~al.}(2019)\citenamefont{Ortiz, Gomes, Morey,
			Winiarski, Bordelon, Mangum, Oswald, Rodriguez-Rivera, Neilson, Wilson
			et~al.}}]{PhysRevMaterials.3.094407}
	\bibinfo{author}{\bibfnamefont{B.~R.} \bibnamefont{Ortiz}},
	\bibinfo{author}{\bibfnamefont{L.~C.} \bibnamefont{Gomes}},
	\bibinfo{author}{\bibfnamefont{J.~R.} \bibnamefont{Morey}},
	\bibinfo{author}{\bibfnamefont{M.}~\bibnamefont{Winiarski}},
	\bibinfo{author}{\bibfnamefont{M.}~\bibnamefont{Bordelon}},
	\bibinfo{author}{\bibfnamefont{J.~S.} \bibnamefont{Mangum}},
	\bibinfo{author}{\bibfnamefont{I.~W.~H.} \bibnamefont{Oswald}},
	\bibinfo{author}{\bibfnamefont{J.~A.} \bibnamefont{Rodriguez-Rivera}},
	\bibinfo{author}{\bibfnamefont{J.~R.} \bibnamefont{Neilson}},
	\bibinfo{author}{\bibfnamefont{S.~D.} \bibnamefont{Wilson}},
	\bibnamefont{et~al.}, \bibinfo{journal}{Phys. Rev. Materials}
	\textbf{\bibinfo{volume}{3}}, \bibinfo{pages}{094407} (\bibinfo{year}{2019}),
	\urlprefix\url{https://link.aps.org/doi/10.1103/PhysRevMaterials.3.094407}.
	
	\bibitem[{\citenamefont{Jiang et~al.}(2021)\citenamefont{Jiang, Yin, Denner,
			Shumiya, Ortiz, Xu, Guguchia, He, Hossain, Liu et~al.}}]{Jiang2021}
	\bibinfo{author}{\bibfnamefont{Y.-X.} \bibnamefont{Jiang}},
	\bibinfo{author}{\bibfnamefont{J.-X.} \bibnamefont{Yin}},
	\bibinfo{author}{\bibfnamefont{M.~M.} \bibnamefont{Denner}},
	\bibinfo{author}{\bibfnamefont{N.}~\bibnamefont{Shumiya}},
	\bibinfo{author}{\bibfnamefont{B.~R.} \bibnamefont{Ortiz}},
	\bibinfo{author}{\bibfnamefont{G.}~\bibnamefont{Xu}},
	\bibinfo{author}{\bibfnamefont{Z.}~\bibnamefont{Guguchia}},
	\bibinfo{author}{\bibfnamefont{J.}~\bibnamefont{He}},
	\bibinfo{author}{\bibfnamefont{M.~S.} \bibnamefont{Hossain}},
	\bibinfo{author}{\bibfnamefont{X.}~\bibnamefont{Liu}}, \bibnamefont{et~al.},
	\bibinfo{journal}{Nature Materials}  (\bibinfo{year}{2021}).
	
	\bibitem[{\citenamefont{Kiesel and Thomale}(2012)}]{PhysRevB.86.121105}
	\bibinfo{author}{\bibfnamefont{M.~L.} \bibnamefont{Kiesel}} \bibnamefont{and}
	\bibinfo{author}{\bibfnamefont{R.}~\bibnamefont{Thomale}},
	\bibinfo{journal}{Phys. Rev. B} \textbf{\bibinfo{volume}{86}},
	\bibinfo{pages}{121105} (\bibinfo{year}{2012}),
	\urlprefix\url{https://link.aps.org/doi/10.1103/PhysRevB.86.121105}.
	
	\bibitem[{\citenamefont{Kiesel et~al.}(2013)\citenamefont{Kiesel, Platt, and
			Thomale}}]{Thomale2013}
	\bibinfo{author}{\bibfnamefont{M.~L.} \bibnamefont{Kiesel}},
	\bibinfo{author}{\bibfnamefont{C.}~\bibnamefont{Platt}}, \bibnamefont{and}
	\bibinfo{author}{\bibfnamefont{R.}~\bibnamefont{Thomale}},
	\bibinfo{journal}{Phys. Rev. Lett.} \textbf{\bibinfo{volume}{110}},
	\bibinfo{pages}{126405} (\bibinfo{year}{2013}),
	\urlprefix\url{https://link.aps.org/doi/10.1103/PhysRevLett.110.126405}.
	
	\bibitem[{\citenamefont{Wang et~al.}(2013)\citenamefont{Wang, Li, Xiang, and
			Wang}}]{PhysRevB.87.115135}
	\bibinfo{author}{\bibfnamefont{W.-S.} \bibnamefont{Wang}},
	\bibinfo{author}{\bibfnamefont{Z.-Z.} \bibnamefont{Li}},
	\bibinfo{author}{\bibfnamefont{Y.-Y.} \bibnamefont{Xiang}}, \bibnamefont{and}
	\bibinfo{author}{\bibfnamefont{Q.-H.} \bibnamefont{Wang}},
	\bibinfo{journal}{Phys. Rev. B} \textbf{\bibinfo{volume}{87}},
	\bibinfo{pages}{115135} (\bibinfo{year}{2013}),
	\urlprefix\url{https://link.aps.org/doi/10.1103/PhysRevB.87.115135}.
	
	\bibitem[{\citenamefont{Feng et~al.}(2021)\citenamefont{Feng, Jiang, Wang, and
			Hu}}]{FENG2021}
	\bibinfo{author}{\bibfnamefont{X.}~\bibnamefont{Feng}},
	\bibinfo{author}{\bibfnamefont{K.}~\bibnamefont{Jiang}},
	\bibinfo{author}{\bibfnamefont{Z.}~\bibnamefont{Wang}}, \bibnamefont{and}
	\bibinfo{author}{\bibfnamefont{J.}~\bibnamefont{Hu}},
	\bibinfo{journal}{Science Bulletin}  (\bibinfo{year}{2021}), ISSN
	\bibinfo{issn}{2095-9273},
	\urlprefix\url{https://www.sciencedirect.com/science/article/pii/S2095927321003224}.
	
	\bibitem[{\citenamefont{Denner et~al.}(2021)\citenamefont{Denner, Thomale, and
			Neupert}}]{denner2021analysis}
	\bibinfo{author}{\bibfnamefont{M.~M.} \bibnamefont{Denner}},
	\bibinfo{author}{\bibfnamefont{R.}~\bibnamefont{Thomale}}, \bibnamefont{and}
	\bibinfo{author}{\bibfnamefont{T.}~\bibnamefont{Neupert}},
	\emph{\bibinfo{title}{Analysis of charge order in the kagome metal
			$a$v$_3$sb$_5$ ($a=$k,rb,cs)}} (\bibinfo{year}{2021}), \eprint{2103.14045}.
	
	\bibitem[{\citenamefont{Park et~al.}(2021)\citenamefont{Park, Ye, and
			Balents}}]{PhysRevB.104.035142}
	\bibinfo{author}{\bibfnamefont{T.}~\bibnamefont{Park}},
	\bibinfo{author}{\bibfnamefont{M.}~\bibnamefont{Ye}}, \bibnamefont{and}
	\bibinfo{author}{\bibfnamefont{L.}~\bibnamefont{Balents}},
	\bibinfo{journal}{Phys. Rev. B} \textbf{\bibinfo{volume}{104}},
	\bibinfo{pages}{035142} (\bibinfo{year}{2021}),
	\urlprefix\url{https://link.aps.org/doi/10.1103/PhysRevB.104.035142}.
	
	\bibitem[{\citenamefont{Lin and Nandkishore}(2021)}]{PhysRevB.104.045122}
	\bibinfo{author}{\bibfnamefont{Y.-P.} \bibnamefont{Lin}} \bibnamefont{and}
	\bibinfo{author}{\bibfnamefont{R.~M.} \bibnamefont{Nandkishore}},
	\bibinfo{journal}{Phys. Rev. B} \textbf{\bibinfo{volume}{104}},
	\bibinfo{pages}{045122} (\bibinfo{year}{2021}),
	\urlprefix\url{https://link.aps.org/doi/10.1103/PhysRevB.104.045122}.
	
	\bibitem[{\citenamefont{Wu et~al.}(2021)\citenamefont{Wu, Schwemmer, Müller,
			Consiglio, Sangiovanni, Sante, Iqbal, Hanke, Schnyder, Denner
			et~al.}}]{wu2021nature}
	\bibinfo{author}{\bibfnamefont{X.}~\bibnamefont{Wu}},
	\bibinfo{author}{\bibfnamefont{T.}~\bibnamefont{Schwemmer}},
	\bibinfo{author}{\bibfnamefont{T.}~\bibnamefont{Müller}},
	\bibinfo{author}{\bibfnamefont{A.}~\bibnamefont{Consiglio}},
	\bibinfo{author}{\bibfnamefont{G.}~\bibnamefont{Sangiovanni}},
	\bibinfo{author}{\bibfnamefont{D.~D.} \bibnamefont{Sante}},
	\bibinfo{author}{\bibfnamefont{Y.}~\bibnamefont{Iqbal}},
	\bibinfo{author}{\bibfnamefont{W.}~\bibnamefont{Hanke}},
	\bibinfo{author}{\bibfnamefont{A.~P.} \bibnamefont{Schnyder}},
	\bibinfo{author}{\bibfnamefont{M.~M.} \bibnamefont{Denner}},
	\bibnamefont{et~al.}, \emph{\bibinfo{title}{Nature of unconventional pairing
			in the kagome superconductors av$_3$sb$_5$}} (\bibinfo{year}{2021}),
	\eprint{2104.05671}.
	
	\bibitem[{\citenamefont{Mazin et~al.}(2014)\citenamefont{Mazin, Jeschke,
			Lechermann, Lee, Fink, Thomale, and Valenti}}]{mazin}
	\bibinfo{author}{\bibfnamefont{I.~I.} \bibnamefont{Mazin}},
	\bibinfo{author}{\bibfnamefont{H.~O.} \bibnamefont{Jeschke}},
	\bibinfo{author}{\bibfnamefont{F.}~\bibnamefont{Lechermann}},
	\bibinfo{author}{\bibfnamefont{H.}~\bibnamefont{Lee}},
	\bibinfo{author}{\bibfnamefont{M.}~\bibnamefont{Fink}},
	\bibinfo{author}{\bibfnamefont{R.}~\bibnamefont{Thomale}}, \bibnamefont{and}
	\bibinfo{author}{\bibfnamefont{R.}~\bibnamefont{Valenti}},
	\bibinfo{journal}{Nature Communications} \textbf{\bibinfo{volume}{5}},
	\bibinfo{pages}{4261} (\bibinfo{year}{2014}).
	
	\bibitem[{\citenamefont{Di~Sante et~al.}(2020)\citenamefont{Di~Sante,
			Erdmenger, Greiter, Matthaiakakis, Meyer, Fernández, Thomale, van Loon, and
			Wehling}}]{DiSante2020}
	\bibinfo{author}{\bibfnamefont{D.}~\bibnamefont{Di~Sante}},
	\bibinfo{author}{\bibfnamefont{J.}~\bibnamefont{Erdmenger}},
	\bibinfo{author}{\bibfnamefont{M.}~\bibnamefont{Greiter}},
	\bibinfo{author}{\bibfnamefont{I.}~\bibnamefont{Matthaiakakis}},
	\bibinfo{author}{\bibfnamefont{R.}~\bibnamefont{Meyer}},
	\bibinfo{author}{\bibfnamefont{D.~R.} \bibnamefont{Fernández}},
	\bibinfo{author}{\bibfnamefont{R.}~\bibnamefont{Thomale}},
	\bibinfo{author}{\bibfnamefont{E.}~\bibnamefont{van Loon}}, \bibnamefont{and}
	\bibinfo{author}{\bibfnamefont{T.}~\bibnamefont{Wehling}},
	\bibinfo{journal}{Nature Communications} \textbf{\bibinfo{volume}{11}}
	(\bibinfo{year}{2020}).
	
	\bibitem[{\citenamefont{Indergand et~al.}(2006)\citenamefont{Indergand,
			L\"auchli, Capponi, and Sigrist}}]{Indergand2006}
	\bibinfo{author}{\bibfnamefont{M.}~\bibnamefont{Indergand}},
	\bibinfo{author}{\bibfnamefont{A.}~\bibnamefont{L\"auchli}},
	\bibinfo{author}{\bibfnamefont{S.}~\bibnamefont{Capponi}}, \bibnamefont{and}
	\bibinfo{author}{\bibfnamefont{M.}~\bibnamefont{Sigrist}},
	\bibinfo{journal}{Phys. Rev. B} \textbf{\bibinfo{volume}{74}},
	\bibinfo{pages}{064429} (\bibinfo{year}{2006}),
	\urlprefix\url{https://link.aps.org/doi/10.1103/PhysRevB.74.064429}.
	
	\bibitem[{\citenamefont{Guo and Franz}(2009)}]{Guo2009}
	\bibinfo{author}{\bibfnamefont{H.-M.} \bibnamefont{Guo}} \bibnamefont{and}
	\bibinfo{author}{\bibfnamefont{M.}~\bibnamefont{Franz}},
	\bibinfo{journal}{Phys. Rev. B} \textbf{\bibinfo{volume}{80}},
	\bibinfo{pages}{113102} (\bibinfo{year}{2009}),
	\urlprefix\url{https://link.aps.org/doi/10.1103/PhysRevB.80.113102}.
	
\end{thebibliography}

\end{document}